\documentclass[12pt]{article}
\usepackage{fullpage,amsmath,amssymb,amsfonts,graphicx,natbib,xcolor,url,placeins,float,mathtools}

\begin{document}
\bibliographystyle{harvard}
\def\covid{\textsc{covid}}
\thispagestyle{empty}
\setcounter{page}{0}
\let\Large\large
\title{{\large{\bf Modeling  Macroeconomic Variations after COVID-19}}}
\author{Serena Ng \bigskip \thanks{Department of Economics, Columbia University and NBER.  420 W. 118 St., New York, NY 10025, email: serena.ng at columbia.edu \newline The author thanks Sai Ma for many helpful discussions. Financial Support from the National Science Foundation (SES  2018368) is gratefully acknowledged. } \\ Columbia University and NBER  }
\date{\today}
\maketitle
\begin{abstract}
The coronavirus  is a global event of historical proportions and just a few months changed the  time series properties of the data in  ways that make many pre-\covid\ forecasting models inadequate. It also  creates a new problem for  estimation of  economic factors and dynamic causal effects because  the variations around the outbreak can be  interpreted as outliers, as shifts to the distribution of existing shocks, or as addition of new shocks.   I take the latter view and use \covid\ indicators  as controls  to 'de-covid' the data prior to estimation. I  find that  economic uncertainty remains high at the end of 2020 even though real economic activity has recovered and \covid\ uncertainty has receded.  Dynamic responses of  variables   to  shocks in a VAR  similar in magnitude and shape to the ones identified before 2020 can be recovered by directly or indirectly   modeling  \covid\ and treating it as exogenous.  These responses to economic shocks are distinctly different from those to a \covid\ shock which are much larger but shorter lived. Disentangling  the two types of shocks can be important in macroeconomic modeling  post-\covid.
\end{abstract}


\noindent JOE Classification: E0, E6

\thispagestyle{empty}
\setcounter{page}{0}
\baselineskip=18.0pt

\newpage
\section{Introduction}
The 2019 coronavirus  (hereafter \covid) is a once in a century global pandemic and remains very much active as of the first draft of this writing, one year after it surfaced in early 2020. It is a health disaster that has resulted in a loss of over half a million lives. It has also disrupted economic activities to such an extent that  the three trillion dollar relief package passed by the Congress was not  enough to offset the economic disruptions \covid\ has caused.   \citet{lmn3-PP:21}  suggests  that \covid\ constitutes  a 192$\sigma$ costly disaster shock.  

In addition to  social and economic disruptions,  \covid\  has also created challenges for the modeling of economic time series.  If we plot a randomly chosen series used in business cycle analysis, we will  likely  see a  spike around March 2020  so large as to  drawf  five decades of  observations preceding it. Without any adjustment, the post-covid\ observations  will dominate  to yield uninterpretable estimates, messing up the   pre-covid fit. A case in point is factor estimation which is used in a variety of economic analysis. Without any adjustment, the first factor known to  load heavily on real activity variables would be 19 standard deviations away from the mean of zero in March 2020 when by way of comparison, the financial crisis in 2008  registered four standard deviations.  

There is no simple solution as there were only two pandemics in the post World War II era, and  no lockdown was enforced in  1957-58 or 1968-69.  As the historical data provide little guide to help understand  the economic implications of global health shocks, how to econometrically handle pandemics  is very much  an open question.  We may treat  these  irregular data points as temporary, but standard outlier adjustments would still leave the real activity factor 13 standard deviations below mean. It could be argued that the extreme values are not void  of economic content and  should not  be 'dummied out'.  Though  treating  the extreme observations as  resulting from  shifts to the underlying distributions may seem appealing,  there are not enough post-\covid\ data to model the instabilities adequately. Introducing restrictions and information may help. \citet{fms:20} corrects  post \covid\ forecasts using information from the 2008 financial crisis.  \citet{primiceri-tambalotti:20}  assumes that \covid\ is a one-period shock that propagates differently from  a typical macroeconomic shock, but whose trajectory can be approximated by a polynomial.  Dynamic responses are then obtained by   calibrating the polynomial to  represent,   for example,  the scenario that the pandemic will dissipate by the end of 2020.   Others  incorporate  information via priors.   \citet{lenza-primceri:20}  specifies a pareto distributed prior to the variance of the shocks while \citet{hkops:20} estimates an additive regression tree   and uses  flexible priors to deal with the extreme values  during the pandemic. Many of these studies were prepared at the early stage of the pandemic  and it is unclear whether the conclusions would hold up in an extended sample that include the subsequent waves.  

 The approach considered in this note also uses additional information, but   I   take as starting point  that  \covid\ is not an economic shock, but  rather a large and  persistent  health event with pervasive economic consequences.   Under this view, the variations in the post-\covid\ economic data are large not because of changes in distribution of variables already in the economic model,   but because the economic  data are no longer  driven by economic shocks alone.  The presence of a new, non-economic shock has  implications   for factor estimation  as the principal components are no longer linear combinations of  economic  variations alone, for  diffusion index forecasting as  new predictors might be relevant, and for estimation of the dynamic causal effects of economic shocks as \covid\ now becomes a confounder. 


To address these issues, I use  \covid\ indicators  either  as controls in regressions to  'de-covid' the data so that economic factors and shocks can be identified, or  as additional predictors  to  account for the persistent nature of  \covid. In a nutshell, I use the \covid\ indicators to either remove    or incorporate  additional  information relevant for the task.  The \covid\ indicators also allow the trajectory of \covid\ to  be determined by the data. Three measures of \covid\ indicators are considered:  hospitalization ($\mathcal H$),  positive cases ($\mathcal P$), and  deaths $(\mathcal D$).   They enter the  de-covid regressions in four ways  reflecting  different assumptions  about  March and April of 2020.    I study their implications in the context of updating  the JLN measure of  economic uncertainty developed in \citet{jurado/ludvigson/ng:13}  since the exercise has a factor estimation step and a forecasting step. I also explore the impact of \covid\ for VAR modeling. In my set up, the issue \covid\ creates is that a $n$ variable VAR now has \covid\ as the  $n+1$-th  shock.

The main findings can be summarized as follows.    All four models find   economic  uncertainty in March/April of 2020 to be at a historical high but there is no corresponding decline in the level of real activity after controlling for \covid.  A decomposition finds that while \covid\ uncertainty has subsided and real activity rebounded by the end of the year,   economic uncertainty even after controlling for \covid\ remains high.  But while the  uncertainty estimates are qualitative similar across methods, the impulse response functions are strongly affected by the de-\covid\ method used.  Dynamic responses to economic shocks  similar  to the ones identified pre-\covid\  can be obtained if  we directly or indirectly  remove the \covid\ variations prior to VAR estimation,  essentially assuming that  \covid\ is exogenous. These responses to economic shocks are distinctly different from those to a \covid\ shock, reinforcing the need to distinguish the two types of shocks in post-\covid\ estimation.  The data  support  the exogeneity assumption.

\section{ Estimation of Common Factors }
  The JLN concept of uncertainty is based on the premise  that uncertainty arises because of lack of predictability with respect to information available, and macroeconomic uncertainty occurs only  when the lack of  economic predictability is  board based.  Three ingredients are needed to make the uncertainty measure operational: factor estimation, forecasting, and volatility estimation.\footnote{In practice, $r_m=8$ economic factors $F_m$ are estimated from a large panel of $N_m$ macro economic variables $X_M$, and $r_f=4$ financial factors $F_f$ from $N_f$ financial time series $X_F$.  In implementation, data for $X_M$ are  taken from  \textsc{fred-md}   and transformed as documented in \citet{fred-md}, and  $X_F$ are based on those used in \citet{ludvigson-ng-jfe}.  The data are  demeaned and standardized prior to factor estimation.}      \citet{lmn4}  makes mean and standard deviation adjustments  prior to factor estimation and add \covid\ variables as predictors.  The procedure   worked well for the August release of FRED-MD  but the adjustments treat too much of the subsequent variations as predictable, making uncertainty counter-intuitively low.  \citet{moran-etal:20} uses the same approach  to update a Q2 measure of Canadian uncertainty. Quarterly data are less affected by  spikes created by \covid\  because  the month-to-month variations tend to average out, so the adjustments may perhaps not be necessary. My focus in what follows  is the  modeling of monthly data.

 Generically let  $X$ be a panel of data with $N$ columns. Let $T_o=720$ be the size of the pre-\covid\ sample running from 1960:3-2020:2.  The $T_0\times N$ matrix of pre-covid data $X$ are assumed to have  a factor structure
\[ X_{it}-\mu_i=\Lambda_i^\prime F_t+e^X_{it}\]
where  $F_t=(F_{1t},\ldots F_{rt})^\prime $ is a vector of $r$ common economic factors  and $e^X_{it}$ is an idiosyncratic error associated with  variable $i$.  
Under conditions in \citet{bai/ng:02} for example, the principal components of $X$ will consistently estimate $F$ and $\Lambda$  up to a rotation matrix.  In practice,  the data are transformed by taking log and first or second difference, and adjusted for outliers  data prior to estimation. The latter amounts to treating as missing those observations   whose deviations from  median  are ten times larger than the difference between values at the top and bottom 25 percentiles, and imputing the missing values using the EM algorithm. Prior to \covid\, this procedure  affects only a few observations during the financial crisis of 2008.

 Let $T_1=730$ be the size of the full sample spanning 1960:3-2020:12, so observations $T_0+1$ to $T_1$ are post \covid. Instead of modeling \covid\ effects through shifts to $F_t$ or  $e_{it}^X$, I allow for a virus factor  $ V_t$ so that    the factor  representation  of the extended sample is
\[ X_{it}-\mu_i=\Lambda_i^\prime F_t+ \Gamma_i V_t+e_{it}^X\]
\paragraph{Assumption A}
\begin{itemize}
\item[i.]  $V_t=0$ when  $t<=T_0$, and for $t> T_0+1$,  $V_t$ is a persistent process with innovations $v_t\sim (0,\sigma^2_v)$.
\item[ii.]  Let $ F_{t}=\Phi^F(L)u^F_{t}$  where $u^F_{kt}\sim (0,\sigma^2_{F_k})$ and $ e^X_{it}=\Phi^X_i(L)u^X_{it}$ where $u^X_{it}\sim (0,\sigma^2_{X_i})$. The   shocks ($u^F_{t},u^X_t$, $v_t$)  are serially and mutually uncorrelated. 

\end{itemize}

 Assumption (i) states that $V_t$ is non-zero only after March 2020 and allows it to be persistent.\footnote{We can also model the non-zero values for the 1968-69 pandemic, but its health and especially the economic impact were small relative to \covid. See \citet{doshi-08} and article in 'Solving the Mystery of the 1957 and 1968 Flu Pandemics in Bloomberg Opinion, March 11, 2021.}    Assumption A.ii assumes that   the  shocks ($u^F_t, u^X_{t}, v_t)$ are serially and mutually uncorrelated. 
  Now $ V_t$ and $F_t$ are both common factors in the sense that they affect a sufficiently large number of series indexed by  $i$. Unlike the pre-\covid\ data, the  principal components of $X$   over the full sample  will no longer be spanned by the economic factors $F_t$ alone. Furthermore, as noted earlier,  the full sample location and scale of the data will be dominated by the last few observations, creating spurious revisions to estimates obtained before \covid.\footnote{The financial factors are much less impacted by the \covid\ observations. The first financial factor based on unadjusted data  is -4.71,  comparable to -4.73 in October 2008  and -6.15 in  October 1987.}    

I consider estimation of economic factors $F_t$ from 'de-\covid' data
\[ x_{it}=\begin{cases} X_{i,t}-\mu_{it}^0 &\quad t\le T_0\\
  X_{it}-\mu^1_{it}  &\quad t>T_0
\end{cases}
\]
for suitably defined $\mu^0_{it}$ and $\mu^1_{it}$. For the pre-\covid\ sample, one can let $\mu^0_{it}=\mu_i^0$ for all $t\le T_0$, and a consistent estimate of $\mu_i^0$ is the mean of series $i$ over the sample up to and including $T_0$. Estimation of $\mu_{it}^1$ is more delicate as \covid\  is persistent and so  its   trajectory also needs to be specified. Even if $V_t$ was observed, there are fewer  than a year's worth of monthly  data to work with. More problematic is that  $F$ and $V$ are both  latent, both  pervasive, and both persistent. Thus recovering both from the post-\covid\ data would require additional information.  I make use of  \covid\ indicators.\footnote{Data are downloaded from \url{https://covidtracking.com/data/download/national-history.csv}. I use the February 21, 2021 vintage. The last release of data was March 7, 2021.}

\begin{table}[ht]
\caption{Proxies for $ V_t$ and $v_t$}
\label{tbl:table2}
\begin{center}
\begin{tabular}{l|rrr|rrrrr} 
& \multicolumn{3}{c}{$ V$} & \multicolumn{3}{c}{$v_t$} \\ \hline
& $\mathcal H$  & $\mathcal P$  & $\mathcal D$ & $\mathcal H$  & $\mathcal P$ & $\mathcal D$\\ \hline
2020-1 &         0 &         2 &  0 &  -&   - & -\\
2020-2 &         0 &        16 &  5 & -&   - & -\\
2020-3 &     6700 &    196830 &  4326  &  8.809&   9.4175 & 6.762\\
2020-4 &    38399 &    876304 &  55315 & 1.745&   1.493 &2.548 \\
2020-5 &    73150 &    718191 &  41137 & 0.644&  -0.199 & -0.296\\
2020-6 &    31513 &    831681 &  19475 & -0.842&   0.147 & -0.747\\
2020-7 &    63105 &   1900163 &  25249 &  0.694&   0.826  & 0.259\\
2020-8 &    61144 &   1457252 &  30244 & -0.015&  -0.265  &0.180\\
2020-9 &    37446 &   1192663 &  23329 & -0.490&  -0.200  & -0.259\\
2020-10 &   53485 &   1892016 & 23545 &   0.356&   0.461 & 0.009\\
2020-11 &   92675 &   4475990 & 37065 &  0.549&   0.861  & 0.453 \\
2020-12 &   126244 &   6323266 & 77112 &   0.309&   0.346 & 0.732\\
2021-01 &   120837 & 6112572   & 95387 & -0.043 & -0.033 &0.212\\
2021-02 &   61054  & 2374243   & 71058& -0.682 & -0.945 & -0.294\\
\hline
\end{tabular}
\end{center}
\hspace*{1.0in}Note: $\mathcal H$ is HospitalizedIncrease,  $\mathcal P$ is PostiveIncrease and  $\mathcal D$ is DeathIncrease. 
\newline \hspace*{1.0in}  Daily data are  aggregated to monthly.
\newline \hspace*{1.0in} Source: \url{covidtracking.com/data/download/national-history.csv}.
\end{table}

  Table \ref{tbl:table2} shows  the daily data for   $\mathcal H=$ 'hospitalizedIncrease,  $\mathcal P$='positiveIncrease', and $\mathcal D$='deathIncrease'   aggregated to monthly.  The cumulative sum of $\mathcal P$  agrees with 11 million cases  documented for the United States in February 2021, while the cumulative sum of $\mathcal D$ is  336802 in December 2020.   Of the three proxies, $\mathcal D$ tends to lag $\mathcal H$ and $\mathcal P$, while  $\mathcal P$  may overstate the situation because one could be tested positive and yet asymptomatic.  From these   \covid\ indicators, I construct three  versions of $v_t$:
\[ v_t=\log\bigg(\frac{V_t}{  V _{t-1}}\bigg).\]
Notably, the $V$ series trend up throughout 2020 but  the $v$  series are less  persistent.\footnote{The data for $\mathcal H$ vary across source but  the $v$ values are quite similar. Since  there were  zero hospitalizations in February but the March value for  $v_t$ is  crucial,  the calculation assumes $\mathcal H=1$ in February.} While the data show  large increases in  $V$   in the summer and the fall,   it is the extraordinary jumps at the outbreak of the pandemic that is problematic for estimation.

Given my presumption that \covid\ and   economic shocks co-exist,  the first task is  to isolate the (predictable and unpredictable) \covid\ variations.  It is natural to identify  $ v_{T_0+1}$ by assuming ($u^F_{t},u^X_t$)=(0,0) at $t=T_0+1$.\footnote{See, for example,  \citet{primiceri-tambalotti:20} and \citet{chudik-etal:20}.}   But  \covid\ was spreading rapidly in April at the same time when the lockdown  was in place, so different interpretations to April are possible.    I consider   four specifications of the following:
\[X_{it}=d_o + \gamma_i D_t + \beta_{i0}  v_t + \beta_{i1}  v_{t-1}+ \beta_{iq}  v_{t-q}+ x_{it}.\]

\begin{center}
\begin{tabular}{l|l|l|l|ll}
$t$ & Model1* & Model 2 &  Model 3& Model 4 \\ \hline
$D_{t=T_0+1}$  & 1 & 1  & 1 &   -\\
$v_{t=T_0+1}$ & 0 & $v_{T_0+1}$ & $v_{T_0+1}$ & $v_{T_0+1}$ \\
$\beta_{i0}$ & $0$ & $0$ & $\ne 0$ & $\ne 0$\\ 
\hline
\end{tabular}

(*) data adjusted for outlier
\end{center}

In all models, $\beta_{i1},\ldots,\beta_{iq}$ are unrestricted and   differ in  the treatment for  March and April of 2020. Model 1 replaces the outliers  by the pre-\covid\ means. Since the interquartile range is now computed on the full sample, the outliers are concentrated in  the two months in question. 
Models 1, 2, and 3   pick up the jump at the outbreak using a one time dummy $D_t$, essentially modeling March as a pure \covid\ shock. By setting both $v_{T_0+1}$ and $\beta_{i0}$ to zero,  Model 1  allocates  all variations in  April   to economic sources and  is expected to generate  large (economic) residuals for that month.  Models 2, 3, and 4  control the extreme values  using \covid\ data.  Model  2  allows   $v_t$ to enter only with a lag while Model 3 allows for contemporaneous effects.    Model 4  allows for economic shocks in March by simply letting current and past effects of $v_t$ be removed from $X_{it}$ as determined by the regression.  To be clear of what the models imply,  I list the   first few post-pandemic values of the regressors  (from March  to July of 2020) for the $\mathcal P$ version below.

\begin{center}
\begin{tabular}{ccc}
\text{Model 1} & \text{Model 2} \\
$
\begin{pmatrix*}[r]
1.000& 1.000 & 0.000 & 0.000 \\
1.000& 0.000 & 0.000 & 0.000 \\
1.000& 0.000 & 1.493 & 0.000 \\
1.000& 0.000 & -0.199 & 1.493 \\
1.000& 0.000 & 0.147 & -0.199 \\
\end{pmatrix*} $
&  

$\begin{pmatrix*}[r]
1.000& 1.000 & 0.000 & 0.000 \\
1.000& 0.000 & 9.418 & 0.000 \\
1.000& 0.000 & 1.493 & 9.418 \\
1.000& 0.000 & -0.199 & 1.493 \\
1.000& 0.000 & 0.147 & -0.199 \\
\end{pmatrix*}
$
\\
\text{Model 3} & \text{Model 4} \\ 
$
\begin{pmatrix*}[r]
1.000& 1.000 & 9.418 & 0.000 & 0.000 \\
1.000& 0.000 & 1.493 &  9.418 & 9.418 \\
1.000& 0.000 & -0.199 & 1.493 & 0.000 \\
1.000& 0.000 & 0.147 & -0.199 & 1.493 \\
1.000& 0.000 & 0.826 & 0.147 & -0.199 \\
\end{pmatrix*} $  &
$ \begin{pmatrix*}[r]
1.000& 9.418 & 0.000 & 0.000 \\
1.000& 1.493 & 9.418 & 0.000 \\
1.000& -0.199 & 1.493 & 9.418 \\
1.000& 0.147 & -0.199 & 1.493 \\
1.000& 0.826 & 0.147 & -0.199 \\
\end{pmatrix*}
$ 
\end{tabular}
\end{center}
\bigskip

  Estimation of  the model  on  post-\covid\ data  gives the fit, which is $\mu^1_{it}$.\footnote{An alternative is to estimate the model  on the full sample with an additional dummy that equals one after $T_0$.}  The  mean adjustments for March 2020 are  quite similar across methods.  For example, based on Model 4, $\hat \mu_0$   \textsc{napmnoi} (new orders)  changes from 55.440  in February to $\hat \mu_1=$ 42.269 in March, \textsc{napm} from 52.983 to 49.122,  \textsc{cumfns} (capacity utilization) from -0.013 to -3.617,  \textsc{umscent} (consumder sentiment) from -0.10 to -12.68,    \textsc{unrate} (unemployment rate) from 0.01 to 0.969, and 
 \textsc{claims} (unemployment claims) from 0.0 to 2.515,  while  housing variables such as \textsc{permit}  as well as  \textsc{awhman} (average man hours) are much less  affected. These results are representative of all four models. However, the April estimates for Model 1 are quite different from those for Models 2, 3, and 4.  Model 1 gives (53.941, 0.005, -0.102)  for \textsc{napm},  \textsc{payems}.  and  \textsc{claims}, while Model 4 yields much larger changes of (42.081, -0.146,  and 0.651) respectively, with the implication that the adjusted data $x$ for Model 1 will have more  extreme values than the other three models.

\begin{center}
\begin{figure}[ht]
\caption{$\hat\mu^1_{it}$, $\mathcal P$ version:  2020:01-2020:12}
\label{fig:fig0}
\centering Model 1
\includegraphics[width=7.0in,height=3.50in]{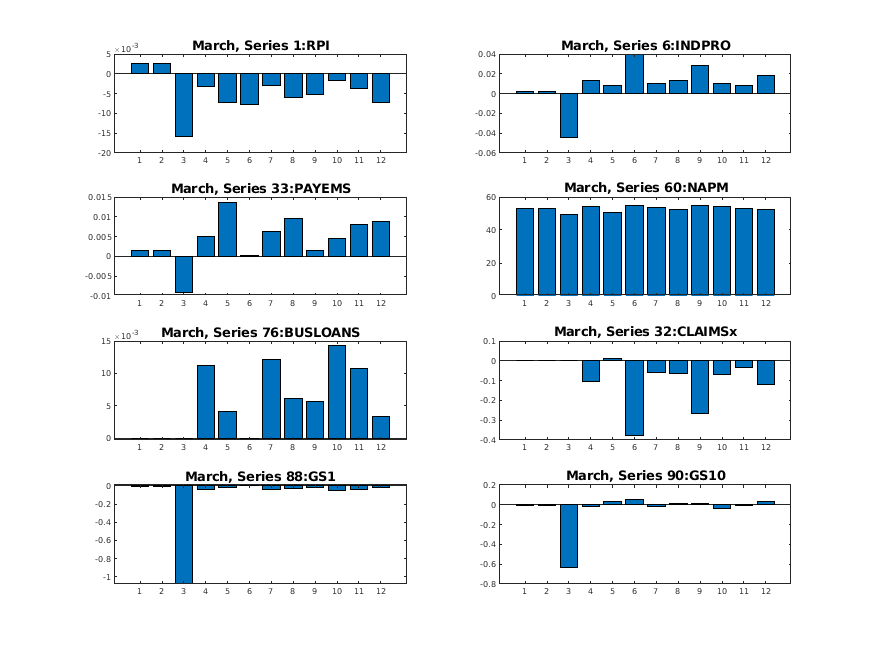}

\centering Model 4
\includegraphics[width=7.0in,height=3.550in]{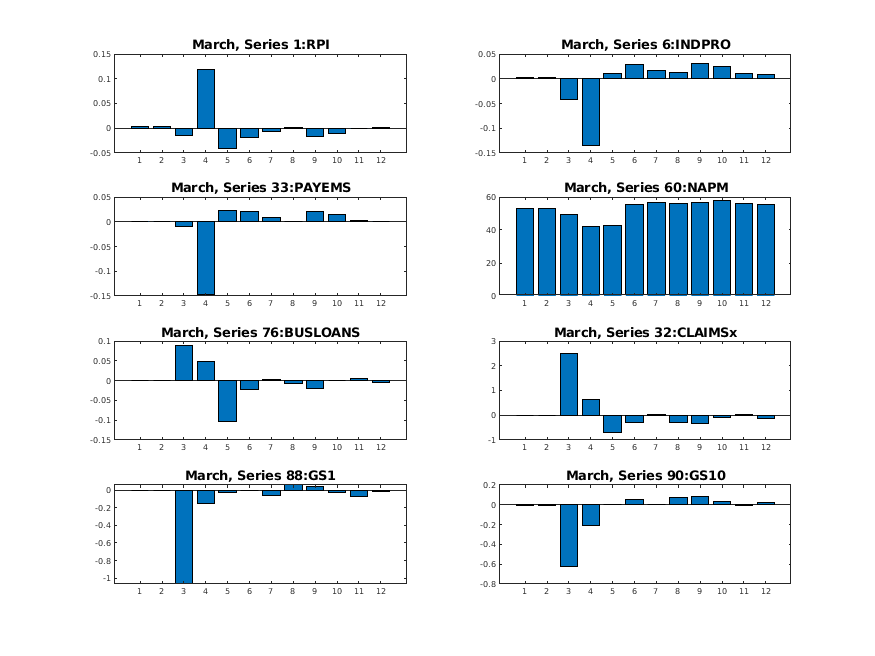}

\end{figure}
\end{center}

 Figure \ref{fig:fig0} plots the 2020 adjustments  for eight selected series.     The impact and subsequent adjustments vary significantly across series and over time. Compared to the Model 1 adjustments  in the top panel,  the Model 4 adjustments in the bottom panel   tend to be more concentrated around April 2020. The differences are most notable for \textsc{rpi}, \textsc{payems}, \textsc{busloans}  and \textsc{claims}.\footnote{The Model 1 adjustments do not include the outlier adjustments.}.  According to  Model 4,   many series are  back to the Janurary/February  levels shortly after April.

 Once $x_{it}=X_{it}-\hat\mu^1_{it}$ is available, I proceed to estimate the factors. Though $x_{it}$   is mean zero in the two respective subsamples, they may not be mean zero when pooled. The stacked data are  demeaned and standardized   prior to factor estimation by the method of principal components.  In the  2020-02 (pre-\covid) vintage of FRED-MD which reports data  up to  2019-12, $F_1$     explains 16\% of the variation in the data. All three post-\covid\ estimates of $F_1$  continue to capture about 16\% of the variations in the extended sample and   are nearly identical up till  2020-02, with pairwise correlations exceeding 0.99. 

\begin{center}
\begin{figure}[ht]
\caption{Factor Estimates: 2019:01-2020:12: Model 4}
\label{fig:F_model4}
\includegraphics[width=7.0in,height=2.750in]{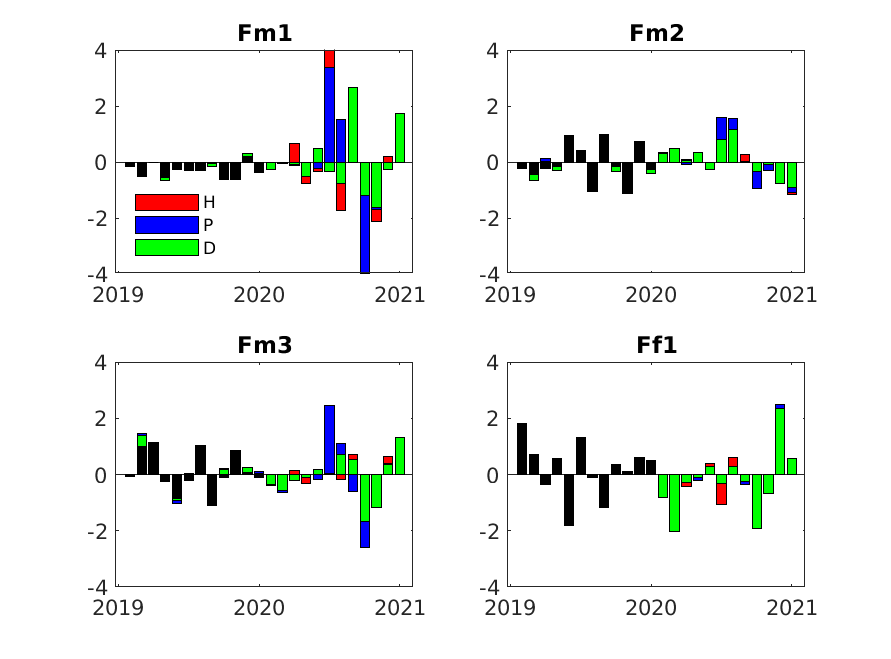}
\end{figure}
\end{center}

Turning to the other factors,  $\hat F_2$ in the pre-\covid\ data loads heavily on term spreads and explains 7\% of variations pre-\covid, while $\hat F_3$ which loads heavily on prices and explains about 6.7\% of the variations. In  the post-\covid\ sample, $\hat F_2$ still loads heavily on term spreads while and $\hat F_3$ still loads heavily on prices. These two factor estimates, along with   $\hat F_ 4, \hat F_7,$ and $\hat F_8$  are nearly perfectly correlated with the pre-\covid\ estimates in the overlapping sample.  However, the $\mathcal H$ and $\mathcal P$  correlations  for  $\hat F_5$ and $\hat F_6$  are  less than 0.8, much lower than the  $\mathcal D$ correlations which are over 0.95. Figure \ref{fig:F_model4} plots the first three macros factors and the first financial  for Model 4. Interestingly, the February values of $\hat F_1, \hat F_2$ and $\hat F_3$  in the de-\covid\ data are more different from  August/September  than  March/April when the lockdown started.

\begin{center}
\begin{figure}[ht]
\caption{Estimates of $F_1$ in 2020:01-2020:12}
\label{fig:F1_acrossmodels}
\includegraphics[width=7.0in,height=3.0in]{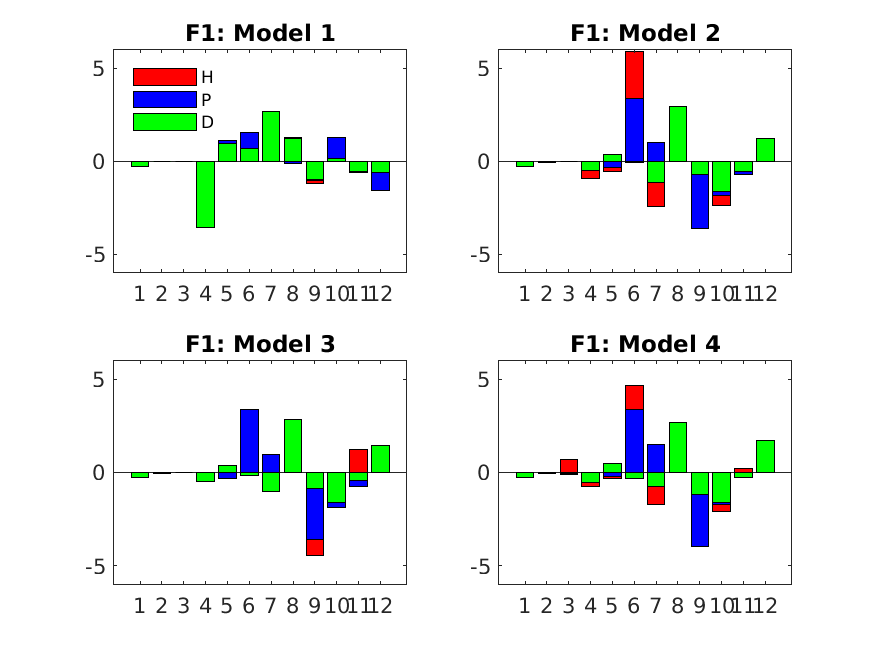}
\end{figure}
\end{center}

Figure \ref{fig:F1_acrossmodels} summarizes the  different estimates  of  $\hat F_1$ in 2020 across models. 
All  suggest a rebound in June presumably due to the stimulus package. 
However, only Model 1 suggests a sharp drop in April with the $\mathcal D$ version being the largest   (-3.59), and ends the year on the negative side.    Models 2, 3, and 4 differ in the magnitude of the rebound in June, but all three estimates of $F_1$ are above the February level at  the end of the year.

\section{Forecasting and Measuring Uncertainty Post-Covid}

The second step of the JLN exercise  involves generating $h$-step ahead prediction errors. Following \citet{stock-watson-diforc}, we form  diffusion index forecasts by  augmenting the estimated factors to an  autoregression:\footnote{For properties of a factor-augmented regression, see \citet{baing-ecta:06,stock-watson-diforc, stock-watson:16}}.
\[ y_{jt+h}=\phi^y_{jh}(L) y_{jt}+\gamma^F_{jh}(L) \hat F_t+\hat \gamma_{jh}^W(L) W_t + v^y_{jt+h}
\] 
Prior to \covid\, $y_{jt}$ is  one of the 134 series in FRED-MD (ie. $X_{jt}$)  after standardization, 
  \[W_t=(\hat F_{mt}^\prime, \hat F_{ft}^\prime,  \hat F_{m,1,t}^2, \hat G_{m,t})^\prime.\]
 where $\hat F_{mt}$  is a set of eight macro factors,   $\hat F_{ft}$ is a set of four  financial factors,   $G_m$ is the first factor in $X_m^2$. A $t$ test with a threshold of 2.56 is used to screen  predictors.

 \covid\ changes this exercise in two ways. First, as  there is  now  a new source of variation in the data,   predictability of $y_{jt+h}$ as measured by  $X_{jt+h}$  must be distinguished from predictability as  measured by $x_{jt+h}$.  Second, the lingering effects of \covid\ on economic activity are not entirely unpredicted after the initial outbreak. \citet{schorfheide-song:20} finds it better to make forecasts shortly after \covid\ using the model estimated pre-\covid, but this cannot be a sustained solution.\footnote{The authors consider a mixed (monthly-quarterly) frequency VAR of eleven variables and finds that for forecasts   made in the end of January, April, and May of 2020,    the model  estimated using data up to the end of 2019  are reasonable  than those based on recursive estimation that includes the post \covid\ data}. I expand the predictor set to include \covid\ indicators: Let
  \begin{equation}
\label{eq2}
W^+_t=(\hat F_{mt}^\prime, \hat F_{ft}^\prime,  \hat F_{m,1,t}^2,  \hat v^ P_t, \hat v_t^D)^\prime.
\end{equation}
Since $y_{jt}$ has been transformed to be stationary, I use $ v_t$ instead of $ V_t$ which is not stationary.  I enter two measures of $ v_t$ in $W_t$  because the number of deaths  tend to lag the number of positive cases, and the two may contain different information.  However,  $G_m$ is no longer a  potential predictor set.  This factor would have to be estimated from  the panel of $X_{jt}^2$ which necessitates additional adjustments.   It is noteworthy that based on the $t$ test criterion, the variables being selected are quite similar before and after \covid.   The lags of  $\hat v_t$ are selected with frequencies ranging from 0.3 to 0.54.

After  an $h$ period ahead diffusion forecast for series  $j$ is obtained from  the factor-augmented regression,  step three of the JLN exercise estimates   stochastic volatility models  for the one-period ahead predictor error $\hat \varepsilon^y_{j,t+1}$ for each series $j$ and  $\hat \varepsilon^F_{k,t+1}$ for each factor $k$.  This volatility estimation step is unaffected by \covid\ though the volatility  estimates will be  higher after \covid. The three-steps yield  $N$ estimates of  individual  uncertainties which  are then  aggregated to form macro-economic uncertainty by equal-weighting.  I will denote the macro-uncertainty measure based on predictability of $x$   by $ U(x)$, to distinguish it from the one based on  predictability   of $X$, which is denoted  $U(X)$.

\begin{figure}[ht]
\caption{One Period Ahead Uncertainty Estimates}
\label{fig:U1}
\includegraphics[width=7in,height=2.50in]{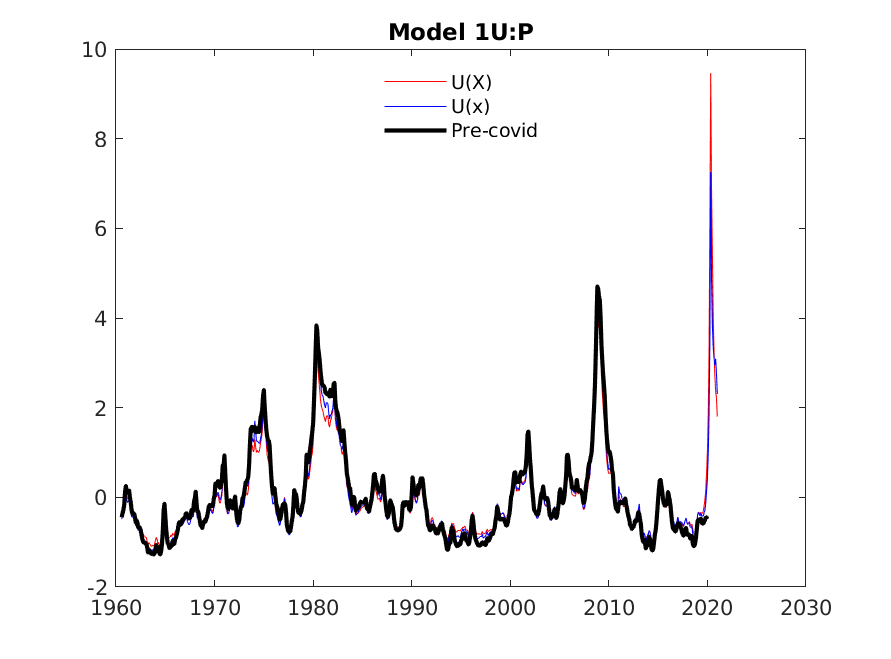}
\includegraphics[width=7in,height=2.50in]{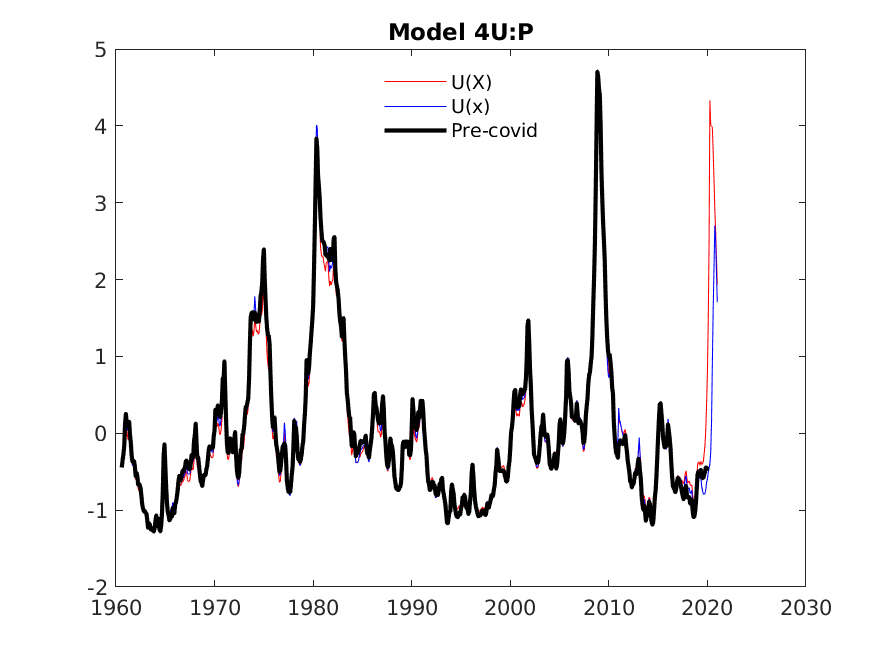}



\includegraphics[width=7in,height=2.50in]{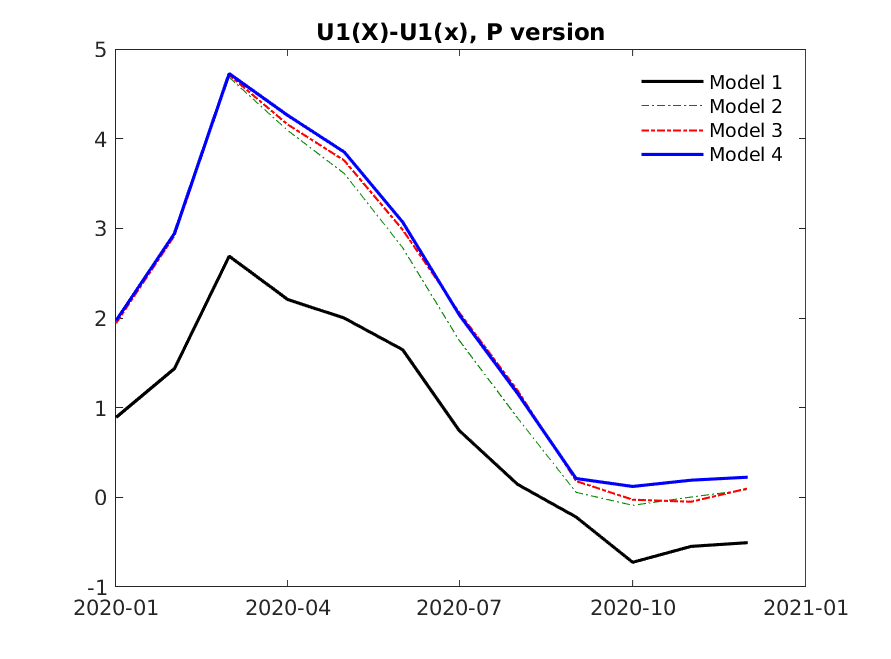}

Note: U(X) is based on predictability of unadjusted data $X$. U(x) is based on predictability of the de-covid data, $x$.

\end{figure}

 Figure \ref{fig:U1} plots the $\mathcal P$ version of uncertainty for Models 1 and 4.   $U(X)$ is  higher than $U(x)$ because $X$ retains the \covid\ variations which contribute to unpredicted forecast errors. Model 1 suggests higher uncertainty in 2020 than Model 4  because of the treatment of the April variations.  But regardless of model and version,
uncertainty is  high in 2020 by historical standard. There are now four episodes of uncertainty that exceed 1.65 standard deviations: the  2007-09,  the 1981-82 and 1973-74 recessions, and \covid.
 
The difference between $U_{M}(X)$ and  $U_M(x)$  can be thought of as 'covid uncertainty'. This is plotted in the bottom panel of Figure \ref{fig:U1} for Model 4.  Though covid uncertainty' is   high in March, it  has subsided by the end of 2020. Also of note is that the  \covid\ uncertainty series for Models 2, 3, and 4 are very similar, suggesting  that   the March dummy  incorporated in Models 2 and 3 is unnecessary, and the agnostic approach used in Model 4 suffices.  The next section shows that Method 4 is also useful in VAR analysis.

\section{Implications for Estimation of Dynamic Causal Effects}
 The data issues created by \covid\ also apply to other forms  of  time series modeling.  Prior to \covid, the model for the $n\times 1$ vector of variables $Y_t$ is
\[ Y_t=\alpha+A(L) Y_{t-1}+B e_t\]
where $e_t$ is  a vector of  $n$ economic shocks. Now  extend the data to include ten months of \covid\ data. Under the view  that  \covid\ is a health shock,  there are now  $n+1$ shocks in the $n$ variable VAR which  cannot be expected to recover $n$ economic shocks.  

\begin{center}
\begin{figure}[ht]
\caption{Dynamic Responses to Unemployment (UR) Shock}
\label{fig:var}
\centering Bivariate VAR, Pre-covid

\includegraphics[width=7.0in,height=1.75in]{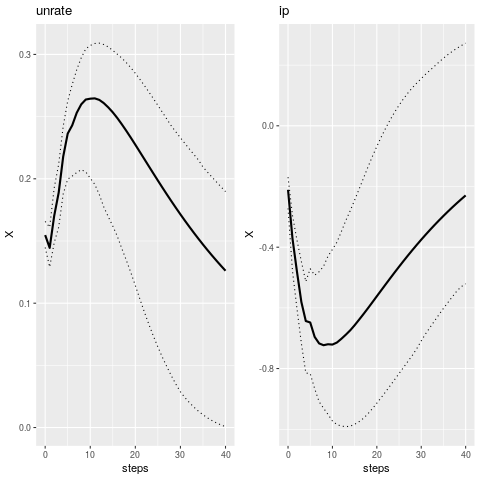}

\centering Bivariate VAR, Post-covid, no adjustment
\includegraphics[width=7.0in,height=1.75in]{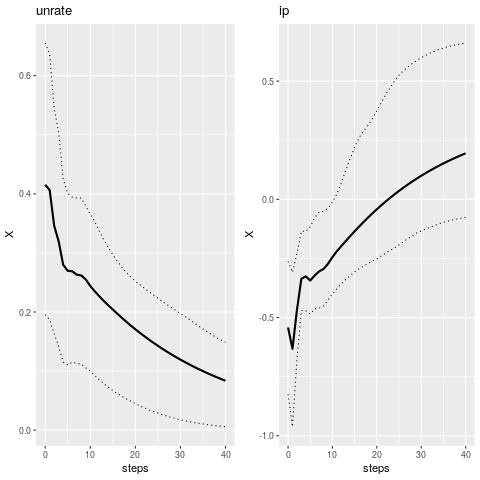}

\centering Bivariate VAR, Post-covid, Model 4
\includegraphics[width=7.0in,height=1.75in]{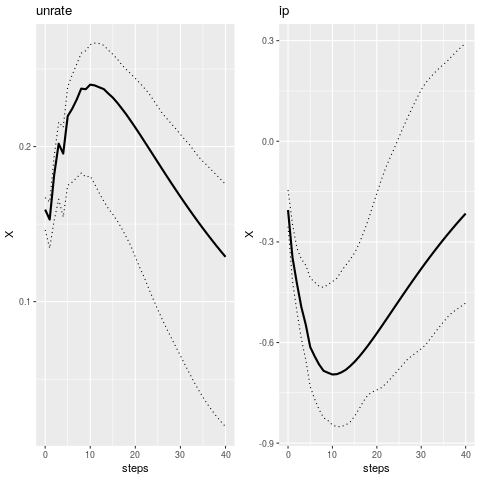}

\centering Three variable VAR
\includegraphics[width=7.0in,height=1.75in]{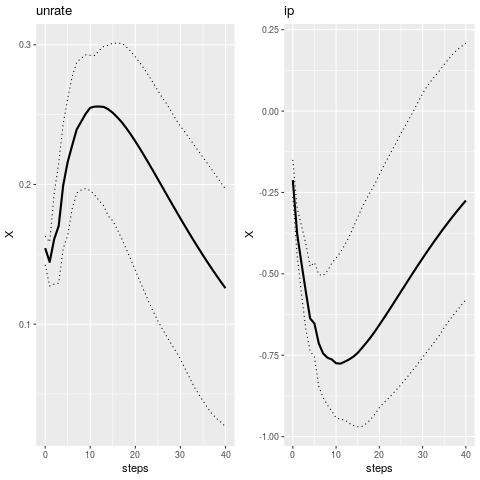}
\end{figure}
\end{center}

To illustrate, consider a VAR(p) in unemplyoment rate (UR) and log industrial production (IP). The top panel of Figure \ref{fig:var} shows  the dynamic responses to a UR shock  using parameters  estimate  over the sample 1961:1-2019:12 with $p=6$ lags. Both responses  have a hump shape shock, peak after eight months, and  are persistent. 
The second panel of  Figure \ref{fig:var} shows that the post-\covid\ responses are generally larger in magnitude  and  distinctively different in shape  from the pre-covid responses in the top panel. In particular, they are no longer hump shaped. Presumably, results like this have prompted \citet{lenza-primceri:20} to scale up the variance of the shocks during the pandemic, and  \citet{ccmm:21} to incorporate outliers in a VAR with stochastic volatility.  Instead of assuming changes to the distributions of economic shocks, I assume that there is  an additional shock.

Following  the arguments above, the idea is to   purge  $v$  from  each of the $n \cdot p$ variables   in the VAR. Note that this is not the same as running a VAR on $n$ de-\covid\ variables.  However, by Frish-Waugh arguments, it is simple to augment the $n$ variable VAR in un-adjusted data $Y$ with $v$ and its lags  as exogenous variables. 
\[ Y_t=\alpha+ \gamma D_t+ \delta  1_{t>T_0} +A(L) Y_{t-1}+\beta(L) v_t+B e_t \quad \tag{VAR-E}\]
where  $D_t$ and each row of $\beta(L)$ are defined as in Models 1 to 4 above. A mean shift dummy is  needed since the de-\covid\ regressions are now run on the full sample.  The role of $v$ and its lags is to partial out the  effects of \covid\ from the dependent variable and the regressors so that  coefficients can be used to construct  dynamic responses  to economic  shocks, holding $v$  and its lags fixed. The third panel of Figure \ref{fig:var} shows   the $\mathcal P$ version of the Model 4  impulse responses based on (VAR-E) along with the 95\% confidence interval. Notably, they  are very similar both in magnitude and in shape to the pre-\covid\ responses in the top panel. The results for $\mathcal D$ and $\mathcal H$  versions are similar. 
  Methods 2 and 3  also retain the hump shaped  responses  but not quite as close the pre-\covid\ estimates as Model 4.  

\begin{center}
\begin{figure}
\caption{Dynamic Responses to Covid Shock}
\label{fig:varcovid}
\centering Three variable VAR
\includegraphics[width=7.0in,height=3.50in]{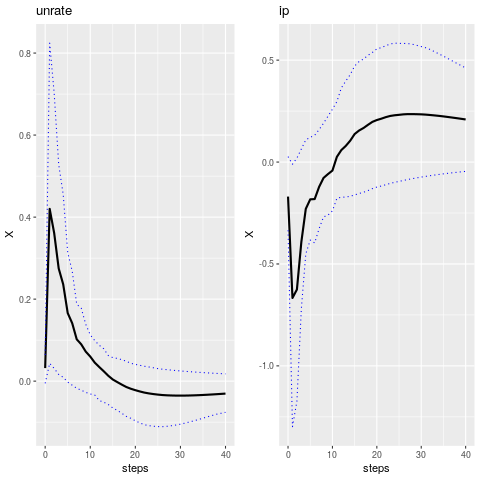}
\end{figure}
\end{center}

It is of interest to consider  a three variable VAR that includes a  \covid\ variable  and ordered first. While  VAR-E  assumes no feedback from UR or IP to $v$, a three variable VAR allows for such feedback. The dynamic responses should be similar if \covid\ is indeed exogenous.\footnote{A six-variable VAR finds likewise that adding the \covid\ indicators as exogenous variables recovers the pre-\covid\ impulse responses. The variables are  unemployment rate (unrate), log of employment (payems), log real consumption of durables (cdur), log real consumption of services (cs), industrial production (ip), and log of the consumption expenditure deflator excluding food and energy.} The bottom panel of Figure \ref{fig:var} verifies that the  responses  of the three variable VAR to a UR shock  are  similar to the ones plotted for VAR-E. However,  these  responses to a UR shock  are distinctly different from the responses    to a \covid\ shock. As seen from Figure \ref{fig:varcovid},  a \covid\ shock has  much larger but shorter lived effects on economic activity. The unemployment rate returns to control after about one year. The effect on IP, while large, is statistically significant for only two months. A two variable VAR without controlling for \covid\ would  confound the responses to the two different shocks, as seen from the second panel of Figure \ref{fig:var}.

\begin{table}[ht]
\caption{Orthogonalized Shocks}
\label{tbl:shocks}
\begin{tabular}{l|r|r|r|r|r|r|r|r|r|r|r|r}
\hline
& \multicolumn{6}{c|}{UNRATE shock } & \multicolumn{6}{c}{IP shock} \\ \hline
  & M0 & M1 & M2 & M3 & M4 & VAR3 & M0 & M1 & M2 & M3 & M4 & VAR3\\

\hline
march & 0.81 & 0.00 & 0.00 & 0.00 & -0.22 & 0.07 & -4.30 & 0.00 & 0.00 & 0.00 & -0.16 & -0.07\\
\hline
april & 9.62 & 19.56 & 0.00 & -0.04 & 0.25 & 0.09 & -11.84 & 0.68 & -0.12 & -0.14 & -0.25 & -0.16\\
\hline
may & -1.28 & -2.43 & 0.34 & 0.09 & 0.14 & 0.04 & 3.22 & 1.40 & -0.04 & -0.20 & 0.01 & -0.10\\
\hline
june & -0.71 & -2.46 & -1.90 & -1.41 & -0.23 & -0.08 & 1.74 & 1.12 & 0.83 & 1.11 & 0.15 & 0.07\\
\hline
cor & 0.83 & 0.81 & 0.96 & 0.97 & 0.97 & 1.00 & 0.95 & 0.96 & 0.97 & 0.98 & 0.98 & 1.00\\
\hline

\end{tabular}

\end{table}
A look at  the orthogonalized shocks  highlights the identification problem that \covid\ creates.
The column labeled M0 is based on post-\covid\ estimation without  controls,  while M1, M2, M2, M4 correspond to the four methods above. These are all bivariate VARs and to be distinguished from  VAR3, which  is  a three-variable  model for  ($v$, UR, IP).  The rows list the values of the identified shocks from March through June as well as the correlation with the corresponding  shock identified pre-\covid\ data over the overlapping sample. Evidently, the  M2, M3, M4  correlations are high and  the  shocks identified for  UNRATE and IP  in April  are small. With no or inadequate adjustments, the M0  and M1 correlations are lower,  and the  shocks identified for UR and IP  in April are  large.  The analysis lends support to treating \covid\ as  exogenous which is the maintained assumption of  Method 4 used  to purge \covid\ effects from the data prior to factor estimation.

Our \covid\ variables are zero before March 2020 which can be approximated by   time series with values  arbitrarily  close to zero except for the few spikes in 2020. Data with such extreme values have heavy tails, and one might wonder if our least squares used to estimate the three variable VAR or the de-\covid\ regressions are valid. Intuitively, the issue is that in a standard regression framework, the response variable should have heavy tails if one of the predictors has heavy tails. Yet, our economic data have fat but not heavy tails. For such problems,   \citet{davis-ng:21} uses a new (heavy-light) framework to dampen the coefficient on the heavy tailed predictor so that it can meaningfully affect the thin-tailed response variable. In that setup, the coefficient on the heavy tailed regressor is consistent at rate $T^{1/\alpha}$ where $\alpha\in(0,2)$ is the tail index, implying that the estimated coefficient on the heavy-tailed regressors are super-consistent.

\section{Conclusion}
  \covid\ can be treated as an outlier,   a health shock  with  macroeconomic consequences, or  an  economic  shock.  I  use  \covid\ indicators to purge the data of their effects  so that  economic  factors can be estimated.  Adding  \covid\ indicators to a VAR as exogenous controls also makes it possible to   recover  impulse responses  to economic shocks similar to the ones estimated  pre-\covid. This note draws attention to the  need to control for \covid\  variations in macroeconomic modeling.

\vspace*{-.75in}


\clearpage
\vspace*{-.5in}

Note: $(H,P,D)$ uses hospitalization data, positive cases, and number of deaths as controls respectively.

\clearpage
\bibliography{../../../prog/sydney/ll,../../../prog1/weekly/notes/weekly,metrics,macro,metrics2,factor,consum}

\end{document}